\documentclass[aps,prb,twocolumn,showpacs]{revtex4}
\usepackage{amssymb}
\usepackage{graphicx}
\usepackage{epsfig}
\usepackage{rotating}

\begin{document}

\title{Nonequilibrium scaling explorations on a 2D Z(5)-symmetric model}
\author{Roberto da Silva}
\email{rdasilva@if.ufrgs.br}
\affiliation{Instituto de F\'{\i}sica, Universidade Federal do Rio Grande do Sul, Av.
Bento Gon\c{c}alves, 9500 - CEP 91501-970, Porto Alegre, Rio Grande do Sul,
Brazil}
\author{Henrique A. Fernandes}
\affiliation{2 - Coordena\c{c}\~{a}o de F\'{\i}sica, Universidade Federal de Goi\'{a}s,
Campus Jata\'{\i}, BR 364, km 192, 3800 - CEP 75801-615, Jata\'{\i}, Goi\'{a}%
s, Brazil}
\author{J. R. Drugowich de Fel\'{\i}cio}
\affiliation{3 - Departamento de F\'{\i}sica, Faculdade de Filosofia, Ci\^{e}ncias e
Letras de Ribeir\~{a}o Preto, Universidade de S\~{a}o Paulo, Avenida
Bandeirantes, 3900 - CEP 14040-901, Ribeir\~{a}o Preto, S\~{a}o Paulo, Brazil}

\begin{abstract}
We have investigated the dynamic critical behavior of the two-dimensional
Z(5)-symmetric spin model by using short-time Monte Carlo (MC) simulations.
We have obtained estimates of some critical points in its rich phase diagram
and included, among the usual critical lines the study of first-order (weak)
transition by looking into the order-disorder phase transition. Besides, we
also investigated the soft-disorder phase transition by considering empiric
methods. A study of the behavior of $\beta /\nu z$ along the self-dual
critical line has been performed and special attention has been devoted to
the critical bifurcation point, or FZ (Fateev-Zamolodchikov) point. Firstly,
by using a refinement method and taking into account simulations
out-of-equilibrium, we were able to localize parameters of this point. In a
second part of our study, we turned our attention to the behavior of the
model at the early stage of its time evolution in order to find the dynamic
critical exponent z as well as the static critical exponents $\beta $ and $%
\nu $ of the FZ-point on square lattices. The values of the static critical
exponents and parameters are in good agreement with the exact results, and
the dynamic critical exponent $z\approx 2.28$ very close of the 4-state
Potts model ($z\approx 2.29$).
\end{abstract}

\maketitle

\section{INTRODUCTION}

In Statistical Mechanics, non-trivial models have been extensively studied
after the exact solution of the two dimensional Ising model \cite%
{Onsager1944}. A lot of authors have devoted an extensive use of several
methods to describe the theory of magnetic systems by studying
generalizations of such model, with more complex and richer phase diagrams.
Among these models, one that deserves special attention is the Z($N$) model
whereas, differently of Ising model whose spin variable can assume only two
values, each spin can assume $N$ values and more than one coupling constant
for $N>4$. This leads to more delicate aspects with phase diagram that is
not completely understood yet, even for example, for small values of $N$
such as $N=5$.

The two-dimensional Z($N$) model contains several known systems as
particular cases, for instance, the Ising $(N=2)$ and \textit{XY} $(N=\infty
)$ models, as well as, the $N$-state scalar and vector Potts (clock) models,
and the Ashkin-Teller model $(N=4)$. For $N\leq 4$, the phase diagram
possesses a traditional second-order phase transition, and for $N=\infty $,
it exhibits a Kosterlitz-Thouless type (KT) phase transition \cite%
{Kosterlitz1973}. But, for what $N$ value does this last phase transition
appear? Several works report that the KT phase transition appears at $N=5$ 
\cite{Jose1977,Alcaraz1981,Fateev1982,Alcaraz1987,Bonnier1989}. The Z(5)
model exhibits a rich phase diagram with first-order transitions, including
the 5-state Potts point \cite{Baxter1973}, two second-order transitions of
the Ising type at Fateev-Zamolodchikov (FZ) integrability points \cite%
{Fateev1982}, and two lines of infinite-order transitions (dual to each
other) of the KT type \cite%
{Jose1977,Alcaraz1981,Nijs1985,Bonnier1989,Bonnier1990,Bonnier1991} (see
dashed lines in Fig. \ref{Fig:phase_diagram}). Several works assert that the
FZ points, henceforth named as \textquotedblleft bifurcation
point\textquotedblright , coincide with the points where the KT transitions
are originated \cite%
{Fateev1982,Bonnier1990,Alcaraz1986,Alcaraz1987,Bonnier1991a}.

So, this interesting model and, in special the bifurcation points (for $N=5$%
), deserves further explorations and non-equilibrium analysis can be an
interesting alternative to obtain not only the static critical exponents but
also the dynamical ones which have not yet been obtained in previous
contributions. Moreover, this approach has proved to be efficient in
determining the critical parameters of several models as shown in recent
works (see for example the Refs. \cite%
{SilvaPRE2013,SilvaCPC2013,SilvaPRE2012}).

In this paper, we present results from the study of the critical properties
of the isotropic ferromagnetic two-dimensional spin model with Z(5)
symmetry, hereafter denoted as Z(5) model, by using time-dependent MC
simulations. As we are dealing with a symmetric model, the two bifurcation
points are also symmetric and possess the same set of critical exponents.
Hence, we concentrated in only one of them. Our contributions are divided in
four parts as follows:

\begin{enumerate}
\item We estimated the critical parameters $x_{1}$ and $x_{2}$ of the
bifurcation point \cite{Fateev1982} by using \ a simple refinement method,
in the context of time-dependent MC simulations which searches the best
power law time decay of magnetization, as proposed in Ref. \cite%
{SilvaPRE2012};

\item We obtained the dynamic critical exponent $z$ and the static critical
exponents $\nu $ and $\beta $ of the two independent order parameters of the
model for the bifurcation point;

\item We explored several points on the self-dual line of the model by
estimating the exponents of its two order parameters. We showed that the
exponents are different along this line but respect a peculiar symmetry.
However, for the particular point corresponding to the 5-state Potts model
the critical exponents assume the same value;

\item We also explored and obtained some estimates of weak first-order
points on the self-dual line and other second-order points on the
soft-disorder transition line using an heuristic method, developed in this
paper, that takes into account the second moment of the order parameters.\ 
\end{enumerate}

This article is organized as follows. In the next section we define the
model and briefly discuss some peculiarities of its phase diagram. In
Section \ref{section3} we present some finite size scaling relations in
non-equilibrium spin systems theory and describe the power laws which are
considered in this work to measure the required exponents and parameters. We
also show how to simulate such behaviors via time-dependent Monte Carlo
simulations. Our results are divided in two sections: In Section \ref%
{Section:ResultsI}, we determined estimates of the phase transition points
in the phase diagram by using a non-equilibrium approach and in Section \ref%
{Section:ResultsII} we specifically showed some estimates of critical
exponents along the self-dual line with special attention to the FZ point.
Finally, in Section \ref{section4} we summarize and conclude our work.

\section{THE MODEL AND ITS PHASE DIAGRAM}

\label{section2}

In this article we have studied the dynamic critical behavior of the Z(5)
model by using short-time Monte Carlo simulations. The most general
Hamiltonian of this model is given by 
\begin{eqnarray}
-\beta \mathcal{H} &=&\sum_{\left\langle i,j\right\rangle }k_{1}\left[ \cos
\left( \frac{2\pi }{5}(n_{i}-n_{j})\right) -1\right]  \nonumber \\
&&+k_{2}\left[ \cos \left( \frac{4\pi }{5}(n_{i}-n_{j})\right) -1\right] ,
\end{eqnarray}%
where $\left\langle i,j\right\rangle $ indicates that the spin variables
interact only with their nearest neighbors, $i$ and $j$ label the sites of a
two-dimensional lattice of size $L\times L$, $k_{1}$ and $k_{2}$ are the two
positive coupling constants, and $n_{i}=0,1,2,3,4$ label the degrees of
freedom of each site of the lattice.

\begin{figure}[th]
\begin{center}
\includegraphics[angle=-90, width=1.1\columnwidth]{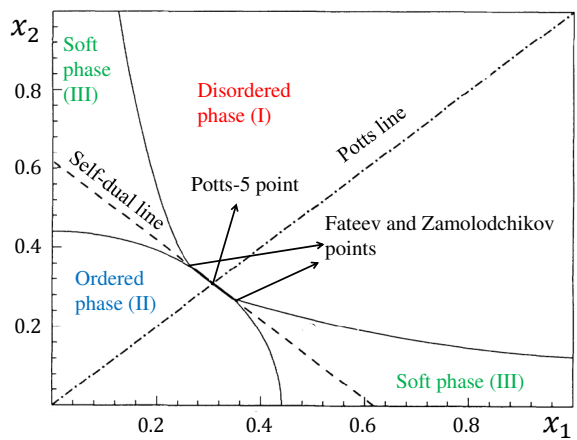}
\end{center}
\caption{(color online) Phase diagram of Z5-model according to the Ref. \protect\cite%
{Rouidi1992}. \textbf{Phase I}: Disordered phase, \textbf{II}: Ordered
phase, and \textbf{III}: Soft phase. The 5-state Potts and FZ points are
specifically indicated on the Self-dual line. The diagram is symmetric with
respect to the Potts physical line. }
\label{Fig:phase_diagram}
\end{figure}

In Fig. \ref{Fig:phase_diagram} (according to Ref. \cite{Rouidi1992}) we can
observe the phase diagram of this model translated to the suitable variables:

\[
x_{1}=\exp \left[ \frac{\sqrt{5}(k_{1}-k_{2})-5(k_{1}+k_{2})}{4}\right] 
\]%
and

\[
x_{2}=\exp \left[ \frac{\sqrt{5}(k_{2}-k_{1})-5(k_{1}+k_{2})}{4}\right] \ 
\text{.} 
\]

In the particular case $k_{2}=k_{1}$ we recover the scalar 5-state Potts
model and for $k_{2}=0$ the clock model. It is interesting to observe that
5-state Potts point corresponds to the meeting between the self-dual line
defined by $x_{1}+x_{2}=(\sqrt{5}-1)/2$ and the Potts physical line $%
x_{1}=x_{2}$, this last one being a symmetry line of the diagram.

In this work we are more concerned with the bifurcation point. Actually, as
can be seen in Fig. \ref{Fig:phase_diagram} the model has two bifurcation
points (FZ points) localized on the self-dual line. The phase transition
line between the FZ points (which includes the Potts point) is of weak first
order, and that on the right(left) of the rightmost(leftmost) FZ point,
there are two continuous transition lines between ordered-soft and
disordered-soft phases. 

However the two bifurcation points are symmetric to each other and have the same set of critical
exponents. For this reason, we took into account only one of them. The ratio
of the coupling constants for the bifurcation point is given by $%
k_{2}/k_{1}=(\sqrt{5}-1)/2\approx 0.618034$. Moreover, there are four order
parameters but only two of them are independent ones \cite{Vanderzande1987},
namely 
\begin{equation}
M_{1}=\left\langle \delta _{n_{i},1}-\delta _{n_{i},2}\right\rangle
\label{M_1}
\end{equation}%
and 
\begin{equation}
M_{2}=\left\langle \delta _{n_{i},1}-\delta _{n_{i},3}\right\rangle \text{,}
\label{M_2}
\end{equation}%
where $\delta _{i,j}$ is the Kronecker's delta.

Since we established the main details of the model in order to calculate the
critical parameters $x_{1}$ and $x_{2}$, as well as the critical exponents $%
z $, $\beta $, and $\nu $, we present in the next section the finite size
scaling developed to describe non-equilibrium spin systems, the
time-dependent power laws obtained from this approach, and some details
about time-dependent MC simulations to be applied.

\section{NON-EQUILIBRIUM DYNAMICS AND TIME-DEPENDENT MC SIMULATIONS}

\label{section3}

Until a few years ago, the numerical calculation of critical exponents was
carried out only in equilibrium. Unfortunately, in this stage, the
measurements of such exponents are very hard due to severe \textit{critical
slowing down} which takes place in the vicinity of the critical temperature.
To circumvent this difficulty, some algorithms were proposed, for instance,
the cluster algorithm \cite{Swendsen1987,Wolff1989} that, although it is
very efficient in the study of static properties, it violates the dynamic
universality class of the specific local dynamics, such as the Model A.

Another way to avoid problems with the \textit{critical slowing down} was
proposed by Janssen, Schaub and Schmittmann \cite{Janssen1989} and Huse \cite%
{Huse1989}, both in 1989. They discovered using renormalization group
techniques and numerical calculations, respectively, that there is
universality and scaling behavior far from equilibrium. Since then, the
so-called short-time regime has become an important method for the study of
phase transitions and critical phenomena.

The dynamic scaling relation obtained by Janssen \textit{et al.} for the 
\textit{k}-th moment of the order parameter, extended to systems of finite
size \cite{Li1995}, is written as 
\begin{equation}
\langle M^{k}\rangle (t,\tau ,L,m_{0})=b^{-k\beta /\nu }\langle M^{k}\rangle
(b^{-z}t,b^{1/\nu }\tau ,b^{-1}L,b^{x_{0}}m_{0}),  \label{eq1}
\end{equation}%
where $t$ is the time evolution, $b$ is an arbitrary spatial rescaling
factor, $\tau =\left( T-T_{c}\right) /T_{c}$ is the reduced temperature and $%
L$ is the linear size of the lattice. The exponents $\beta $ and $\nu $ are
the equilibrium critical exponents associated with the order parameter and
the correlation length, and $z$ is the dynamic exponent characterizing
temporal correlations in equilibrium. Here, the operator $\langle \ldots
\rangle $ denotes averages over different configurations due to different
possible time evolution from each initial condition of a given initial
magnetization $m_{0}$. For a large lattice size $L$ and small initial
magnetization $m_{0}$ at the critical temperature $(\tau =0)$, the Eq. (\ref%
{eq1}) is governed by the new dynamic exponent $\theta $, according to 
\begin{equation}
\langle M\rangle _{m_{0}}\sim m_{0}t^{\theta }\text{,}  \label{theta}
\end{equation}%
if we choose the scaling factor $b=t^{1/z}$. This new exponent characterizes
the so-called \textit{critical initial slip}, the anomalous behavior of the
order parameter when the system is quenched to the critical temperature $%
T_{c}$.

Besides, a new critical exponent $x_{0}$, which represents the anomalous
dimension of the initial magnetization $m_{0}$, is introduced to describe
the dependence of the scaling behavior on the initial conditions. This
exponent is related to $\theta $ as $x_{0}=\theta z+\beta /\nu $. Actually
the relaxation of spin systems is determined by two different behaviors,
this initial slip and a second behavior corresponding to a power-law decay.
This can be derived from the Eq. (\ref{eq1}). After the scaling $b^{-1}L=1$
at the critical temperature $T=$ $T_{c}$, the first ($k=1$) moment of the
order parameter is $\langle M\rangle (t,L,m_{0})=L^{-\beta /\nu }\langle
M\rangle (L^{-z}t,L^{x_{0}}m_{0})$.

Denoting $u=tL^{-z}$ and $w=L^{x_{0}}m_{0}$, one has $\langle M\rangle
(u,w)=L^{-\beta /\nu }\langle M\rangle (L^{-z}t,L^{x_{0}}m_{0})$. The
derivative with respect to $L$ is given by:

\begin{eqnarray*}
\partial _{L}\langle M\rangle &=&(-\beta /\nu )L^{-\beta /\nu -1}\langle
M\rangle (u,w) \\
&&+L^{-\beta /\nu }[\partial _{u}\langle M\rangle \partial _{L}u+\partial
_{w}\langle M\rangle \partial _{L}w]\text{,}
\end{eqnarray*}%
where explicitly we have $\partial _{L}u=-ztL^{-z-1}$ and $\partial
_{L}w=x_{0}m_{0}L^{x_{0}-1}$. In the limit $L\rightarrow \infty $, which
implicates in $\partial _{L}\langle M\rangle \rightarrow 0$, one has $%
x_{0}w\partial _{w}\langle M\rangle -zu\partial _{u}\langle M\rangle -\beta
/\nu \langle M\rangle =0$. The separability of the variables $u$ and $w$,
i.e., $\langle M\rangle (u,w)=M_{u}(u)M_{w}(w)$ leads to%
\[
x_{0}wM_{w}^{\prime }/M_{w}=\beta /\nu +zuM_{u}^{\prime }/M_{u}\text{,} 
\]%
where the prime means the derivative with respect to the argument. Since the
left-hand side of this equation depends only on $w$ and the right-hand side
depends only on $u$, both sides must be equal to a constant $c$. Thus, $%
M_{u}(u)=u^{c/z}-\beta /(\nu z)$ and $M_{w}(w)=w^{c/x_{0}}$, resulting in $%
\left\langle M\right\rangle (u,w)=m_{0}^{c/x_{0}}L^{\beta /\nu }t^{(c-\beta
/\nu )/z}$. Returning to the original variables, one has $\langle M\rangle
(t,L,m_{0})=m_{0}^{c/x_{0}}t^{(c-\beta /\nu )/z}$.

On one hand, by choosing $c=x_{0}$ at criticality ($\tau =0$), one obtains $%
\langle M\rangle _{m_{0}}\sim m_{0}t^{\theta }$, where $\theta =(x_{0}-\beta
/\nu )/z$ that corresponds to a regime of small initial magnetization soon
after a finite time scaling $b=t^{1/z}$ in Eq.~\ref{eq1}. This leads to $%
\left\langle M\right\rangle (t,m_{0})=t^{-\beta /(\nu z)}\langle M\rangle
(1,t^{x_{0}/z}m_{0})$. By calling $x=t^{x_{0}/z}m_{0}$, an expansion of the
averaged magnetization around $x=0$ results in $\langle M\rangle
(1,x)=\langle M\rangle (1,0)+\left. \partial _{x}\langle M\rangle
\right\vert _{x=0}x+\mathcal{O}(x^{2})$. By construction $\langle M\rangle
(1,0)=0$ and, since $u=t^{x_{0}/z}m_{0}\ll 1$, we can discard quadratic
terms resulting in $\langle M\rangle _{m_{0}}\sim m_{0}t^{\theta }$. This
anomalous behavior of initial magnetization is valid only for a
characteristic time scale $t_{\max }$ $\sim m_{0}^{-z/x_{0}}$.

On the other hand, the choice $c=0$ corresponds to the case where the system
does not depend on the initial trace and $m_{0}=1$ leads to simple power
law: 
\begin{equation}
\langle M\rangle _{m_{0}=1}\sim t^{-\beta /(\nu z)}  \label{decay_ferro}
\end{equation}%
that similarly corresponds to the decay of magnetization (for $t>t_{\max }$)
of a system previously evolved from an initial small magnetization $(m_{0})$%
, and that had its magnetization increased according to Eq. \ref{theta} up
to a peak.

For $m_{0}=0$, it is not difficult to show that the second moment of the
magnetization is given by 
\begin{equation}
\left\langle M^{2}\right\rangle _{m_{0}=0}\sim t^{\varsigma }\;,  \label{M2}
\end{equation}%
with $\varsigma =(d-2\beta /\nu )/z$, where $d$ is the dimension of the
system. By using short-time MC simulations, where lattices are suitably
prepared with a fixed initial magnetization, many authors have obtained the
dynamic exponent $z$ as well as the static ones $\beta $ and $\nu $, for
many different models (see, for example, two good reviews can be found in
Refs. \cite{Albano2011}, \cite{Zheng1998}).

In order to estimate independently the critical exponents, we can, firstly,
determine $z$ by using a power law that mixes initial conditions \cite%
{Silva20021} as follows 
\begin{equation}
F_{2}(t)=\frac{\left\langle M^{2}\right\rangle _{m_{0}=0}}{\left\langle
M\right\rangle _{m_{0}=1}^{2}}\sim t^{\xi }\text{,}  \label{z}
\end{equation}%
where $\xi =d/z$. With the estimate of $\xi $, denoted here by $\widehat{\xi 
}$, we are able to obtain an estimate of $z$ (given by $\widehat{z}=d/%
\widehat{\xi }$) independent of other parameters. In order to obtain $\nu $,
we use an alternative power law. When considering $m_{0}=1$ in Eq. \ref{eq1}%
, one can see that there is no dependence on the initial configurations.
Therefore, when $L\rightarrow \infty $, one can $\langle M\rangle (t,\tau
)=b^{-k\beta /\nu }\langle M\rangle (b^{-z}t,b^{1/\nu }\tau )$. By scaling $%
b^{-z}t=1$, we have $\langle M\rangle (t,\tau )=t^{-\beta /(\nu
z)}f(t^{1/(\nu z)}\tau )$ where $f(x)=$ $\langle M\rangle (1,x)$ and so $%
\partial \ln \langle M\rangle (t,\tau )/\partial \tau =\frac{1}{\langle
M\rangle }\frac{\partial }{\partial \tau }\langle M\rangle =t^{1/(\nu
z)}f(t^{1/(\nu z)}\tau )$. Therefore we have 
\begin{equation}
D(t)=\left. \frac{\partial \ln \langle M\rangle }{\partial \tau }\right\vert
_{\tau =0}=f_{0}\cdot t^{1/(\nu z)}\sim t^{\phi }  \label{1niz}
\end{equation}%
where $f_{0}=f(0)$ is a constant and $\phi =$ $1/(\nu z)$. Since we have
already obtained the exponent $z$, we are able to obtain $\nu $. With these
two exponents in hand, we can obtain $\beta $ by estimating the exponent $%
\mu =\beta /(\nu z)$ from Eq. \ref{decay_ferro}.

In order to simulate numerically the theoretical moments of the
magnetization of the spin systems as functions of time, we used a local
dynamic evolution of the spins which are updated by the heat-bath algorithm.
In our simulations we used two different initial states: to obtain the power
laws giving by the Eqs. \ref{decay_ferro} and \ref{1niz}, we used the
initial ordered state, i.e., $m_{0}=1\ $($\sigma _{i}\equiv 1$, $%
i=1,...,N=L^{d}$). On the other hand, when considering the Eq. \ref{M2} we
used a initial state with $m_{0}=0$, i.e., the spins of each site were
chosen at random on the sites but keeping the same proportion -- $L^{d}/5$
spins of each type: $\sigma _{i}=0,1,2,3,4$. Here it is important to mention
that $m_{0}=0$ for any order parameter proposed in our analysis [Eqs. \ref%
{M_1} and \ref{M_2}].

In the context of time-dependent MC simulations, the magnetization ($k=1$)
and its higher moments ($k>1$) have statistical estimators for the
theoretical moments (\ref{eq1}) given by 
\[
\left\langle M^{k}\right\rangle (t)=\frac{1}{N_{run}L^{d}}%
\sum\limits_{j=1}^{N_{run}}\left( \sum\limits_{i=1}^{L^{d}}\sigma
_{i,j}(t)\right) ^{k}\text{,} 
\]%
where $\sigma _{i,j}(t)$ denotes the $i$-th spin variable on the lattice at $%
t$-th MC step of the $j$-th run. Here $N_{run}$ denotes the number of
different repetitions (runs) or different time series used to compute the
averages.

\section{Results I: Exploring the phase diagram via non-equilibrium MC
simulations}

\label{Section:ResultsI}

Our initial plan was to study the phase transition points of the Z(5) model
via time-dependent MC simulations by estimating the best $x_{2}$ given as
input the parameter $x_{1}$ according to the phase diagram (see Fig. \ref%
{Fig:phase_diagram}). We performed this task for several points in this
diagram and the analysis was carried out by using an approach developed in 
\cite{SilvaPRE2012} in the context of generalized statistics. This tool had
also been applied successfully to study multicritical points, for example,
tricritical points \cite{SilvaCPC2013}\cite{SilvaPRE2002} and Lifshitz point
of the ANNNI model \cite{SilvaPRE2013}.

Since at criticality is expected that the order parameter obeys the power
law behavior of Eq. \ref{decay_ferro}, we fixed the value of $x_{1}$ and
changed the value of $x_{2}$ according to a resolution $\Delta x_{2}$. Then,
we calculated the known coefficient of determination \cite{Trivedi2002}
that, for our case, is given by: 
\begin{equation}
r=\frac{\sum\limits_{t=1}^{N_{MC}}(\overline{\ln \langle M\rangle }-a-b\ln
t)^{2}}{\sum\limits_{t=1}^{N_{MC}}(\overline{\ln \left\langle M\right\rangle 
}-\ln \langle M\rangle (t))^{2}}\text{,}  \label{determination_coefficient}
\end{equation}%
with $\overline{\ln \langle M\rangle }=(1/N_{MC})\sum%
\nolimits_{t=1}^{N_{MC}}\ln \langle M\rangle (t)$, for each value $%
x_{2}=x_{2}^{(\min )}+i\Delta x_{2}$, with $i=1,...,n$, where $%
n=\left\lfloor (x_{2}^{(\max )}-x_{2}^{(\min )})/\Delta x_{2}\right\rfloor $%
, and the critical value corresponds to $x_{2}^{(opt)}=\arg \max_{x_{2}\in
\lbrack x_{2}^{(\min )},x_{2}^{(\max )}]}\{r\}$. The coefficient $r$ has a
very simple explanation: it measures the ratio: (expected variation)/(total
variation). The bigger the $r$, the better the linear fit in log-scale, and
therefore, the better the power law which corresponds to the critical
parameter except for an error $O(\Delta x_{2})$.

As we are dealing with a rich phase diagram, a careful analysis of the
order of the phase transition is necessary, mainly when taking into account
first-order \textquotedblleft critical\textquotedblright\ points. As pointed
out earlier, the phase diagram of the Z(5) model possesses two second-order
phase transition points which coincide with the FZ integrability points, as
well as two lines of infinite-order transition (dual to each other) also
known as self-dual lines. The phase transitions of the points on these lines
which extend from the 5-state Potts point to the FZ points are expected to
be of first-order. Although it is not expected a power law behavior of the
order parameter at strong first-order points, it is possible to obtain this
behavior for weak first-order ones, whereas for $k>k_{c}$ a disorder
metastable state vanishes at a certain $k^{\ast }$ and, for $k<k_{c}$, there
is an ordered metastable state which disappears at $k^{\ast \ast }$. Both
parameter values look like critical points if the system remains in the
disordered or ordered metastable states, and so in both points a power law
behavior must be observed as studied by Schulke and Zheng \cite{Schulke2000}
through the analysis of the weakness of first-order phase transition in the $%
q$-state Potts model. In that case a good estimate for $k_{c}$ would be $%
(k^{\ast }+k^{\ast \ast })/2$. For the 5-state Potts model, for example, the
difference between the pseudo critical points $k^{\ast }$ or $k^{\ast \ast }$
and $k_{c}$ is in the fourth decimal digit. Moreover, the difference between
power laws obtained from the pseudo critical points and $k_{c}$ is observed
for $t\sim 1000$ MC steps.

Since the self-dual line of the Z(5) model is analytically described by $%
x_{2}=\frac{(\sqrt{5}-1)}{2}-x_{1}$ and the points extending from $x_{1}=(%
\sqrt{5}-1)/4\approx 0.30901...$ to (but not including) the FZ point (which
corresponds to $x_{1}\approx 0.3473834...$) are points of weak first-order
transition, we determined the corresponding $x_{2}$ via method previously
described. In this case, by looking into the difference between $x_{2}$%
(exact) and $x_{2}$(simulation), it was possible to have a measure of
weakness of the considered points.

\begin{table*}[tbp] \centering%
\begin{tabular}{lllllll}
\hline\hline
$x_{1}$ & $x_{2}$(exact) & $x_{2}^{opt}$(simulation) & $r(x_{2}^{opt}-\Delta
x_{2})$ & $r(x_{2}^{opt})$ & $r(x_{2}^{opt}+\Delta x_{2})$ & $%
(x_{2}^{opt})^{(2)}$ \\ \hline\hline
Potts 5 & $0.30901...$ & $0.308(2)$ & $0.994251$ & $0.999605$ & $0.999557$ & 
$0.3094(1)$ \\ 
$0.31$ & $0.30803...$ & $0.308(2)$ & $0.997386$ & $0.999514$ & $0.998977$ & $%
0.3083(1)$ \\ 
$0.32$ & $0.29803...$ & $0.298(2)$ & $0.997535$ & $0.999696$ & $0.997920$ & $%
0.2979(1)$ \\ 
$0.33$ & $0.28803...$ & $0.288(2)$ & $0.998707$ & $0.999715$ & $0.998626$ & $%
0.2873(1)$ \\ 
$0.34$ & $0.27803...$ & $0.278(2)$ & $0.998385$ & $0.999572$ & $0.998690$ & $%
0.2781(1)$ \\ 
FZ & $0.27065...$ & $0.270(2)$ & $0.999401$ & $0.999701$ & $0.999168$ & $%
0.2702(1)$ \\ \hline\hline
\end{tabular}%
\caption{Analysis of the weak first-order transitions until the critical
point FZ}\label{Table:optmization_weak_first_order}%
\end{table*}%

In TABLE \ref{Table:optmization_weak_first_order}, third column, we show our
results for $x_{2}$ ($x_{2}^{opt}$) for five points along the self-dual line
that whose transitions are expected to be of first-order, as well as for the
FZ point (sixth line). In order to obtain these results, we used resolution
of $\Delta x_{2}=0.002$ and applied a simple algorithm that makes a process
of refinement of the parameter in order to localize the best $x_{2}$ along
the simulations. These values must be compared to the exact predictions of
the self-dual line (second column). It is important to notice that the
columns 4, 5, and 6 represent, respectively, the values of $r$ obtained for
the fits with respective values of $x_{2}$: $x_{2}^{opt}-\Delta x_{2}$, $%
x_{2}^{opt}$, and $x_{2}^{opt}+\Delta x_{2}$. For instance, we observe that,
for the Potts point $r(x_{2}^{opt}-\Delta x_{2})=0.994251$, $r(x_{2}^{opt})=$
$0.999605$ and $r(x_{2}^{opt}+\Delta x_{2})=0.999557$. From that, we applied
a second refinement for the interval $[x_{2}^{opt}-\Delta x_{2},\
x_{2}^{opt}+\Delta x_{2}]$ by using $\Delta x_{2}=10^{-4}$ and we found $%
0.3094(1)$ (seventh column). When compared to the exact value $0.30901...$
we observed an error only in the fourth decimal place which is reasonable
according to lattice used in our MC simulations for this optimization, $%
L=160 $.

Now, since we analyzed the first-order (weak) transition up to the
bifurcation point, we turned our attention to points after it via time
dependent MC simulations. According to these phase diagram (Fig. \ref%
{Fig:phase_diagram}), after the bifurcation point, $x_{1}>0.3473834...$,
there are two second-order lines separating the ordered and disordered
phases and the soft one.

\begin{figure*}[th]
\begin{center}
\includegraphics[width=\columnwidth]{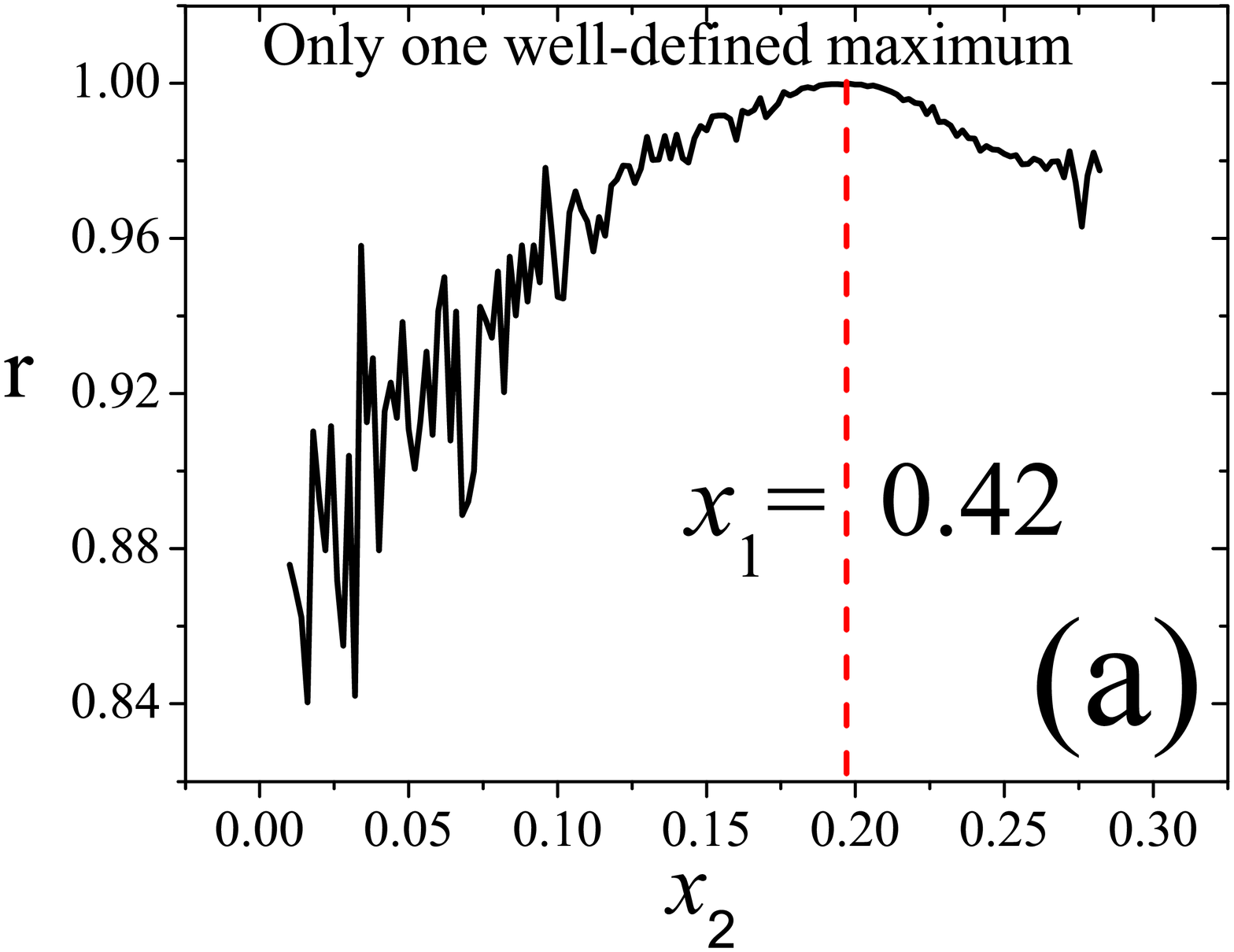} \includegraphics[width=%
\columnwidth]{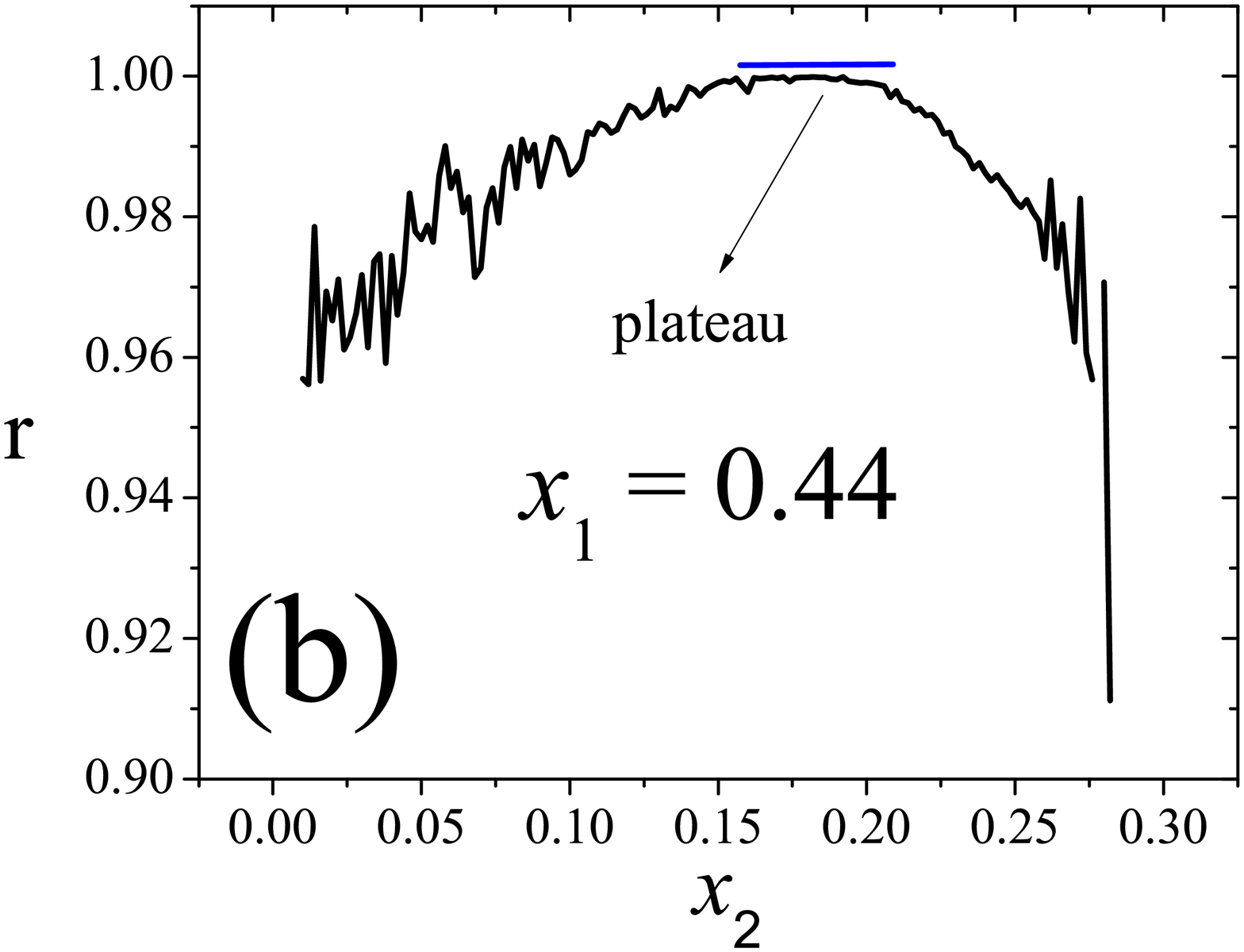} \includegraphics[width=%
\columnwidth]{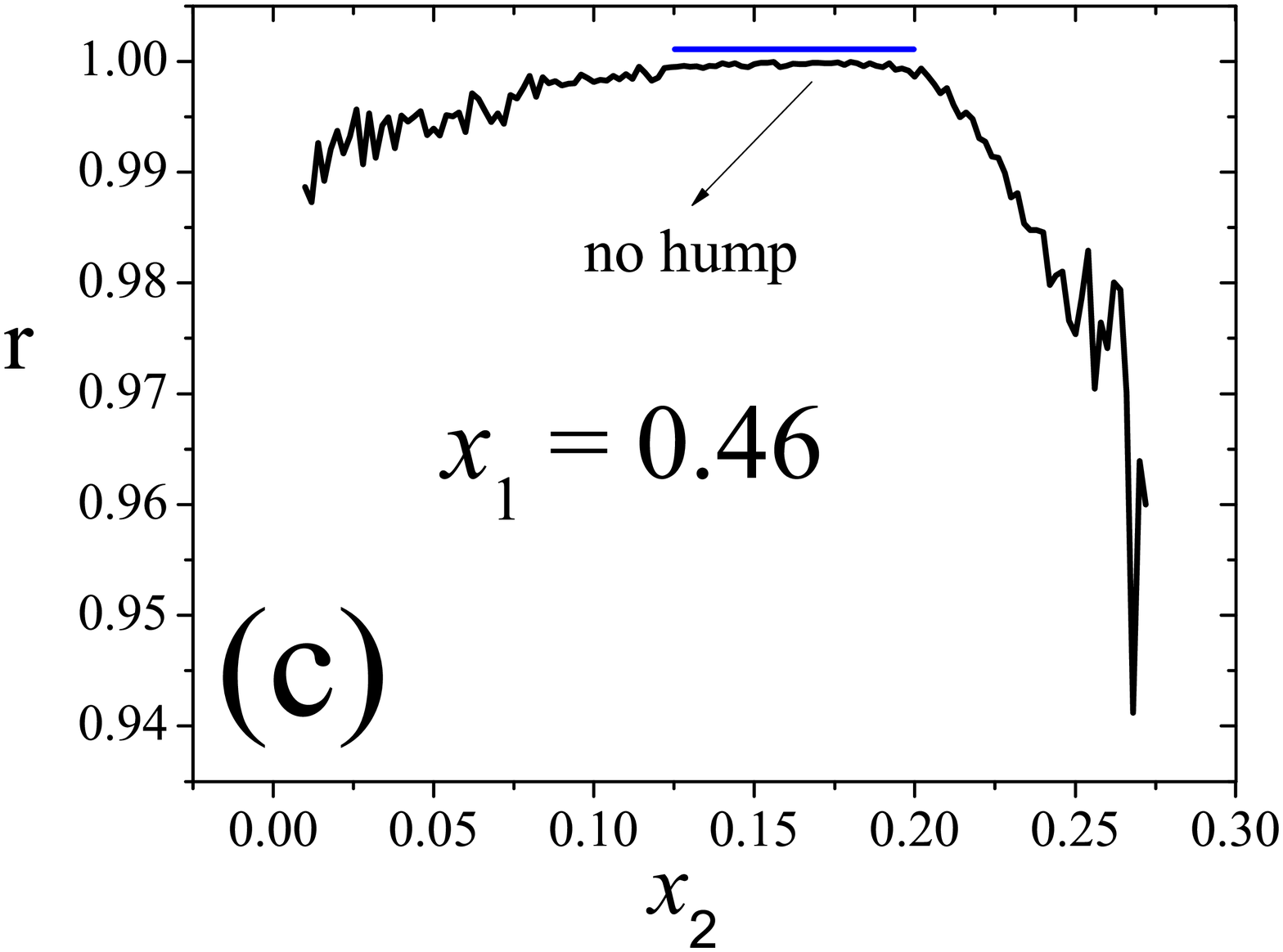} \includegraphics[width=%
\columnwidth]{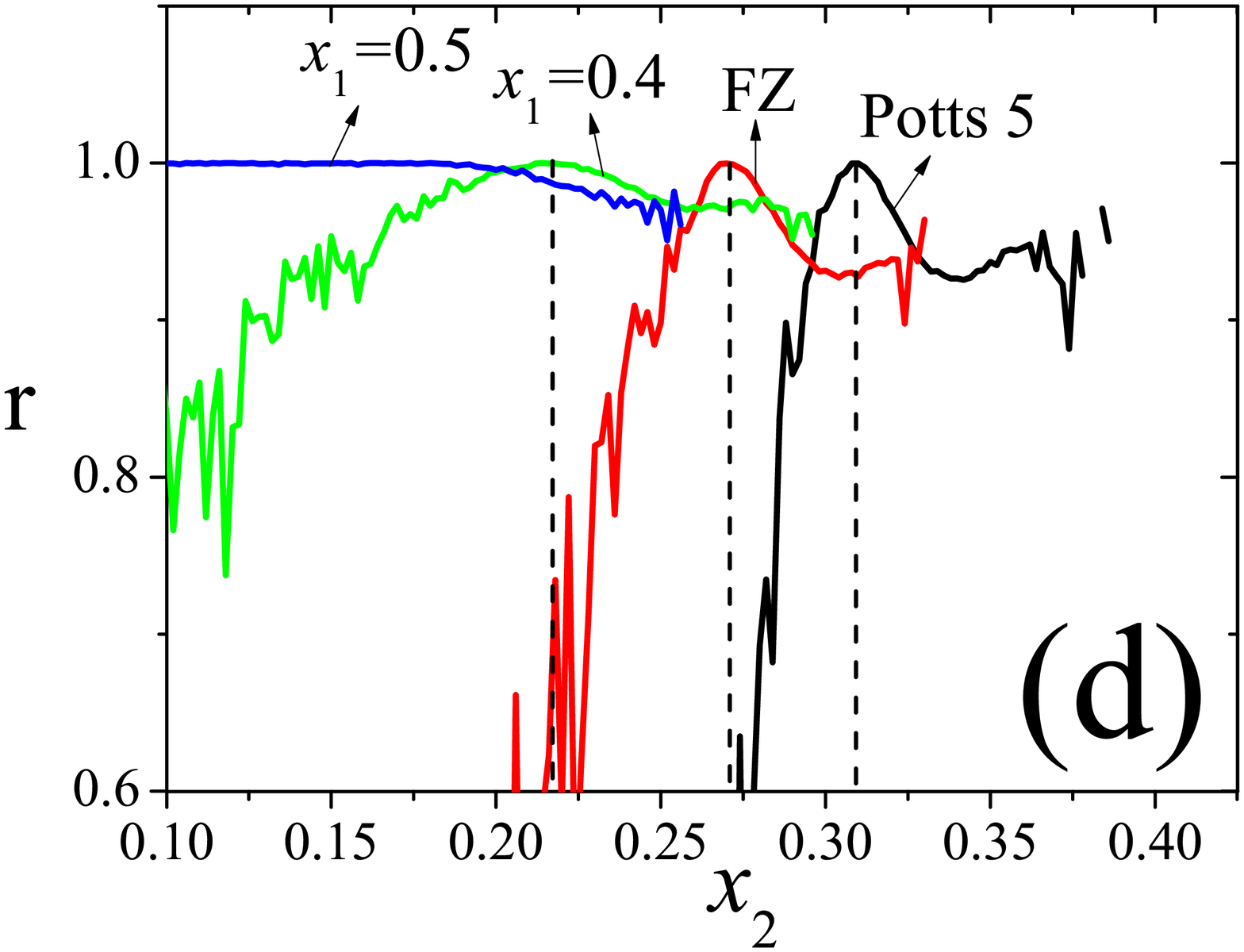}
\end{center}
\caption{(color online) \textbf{Plot (a)}: Refinement process for the input $x_{1}=0.42$. A
clear point where $r$ is maximum is found. The \textbf{plots (b)} and 
\textbf{(c)} show, respectively, the refinement for $x_{1}=0.44$ and $%
x_{1}=0.46$. In these cases there is no a notorious optimization point since 
$x_{1}=0.44$ is the last point where we expect to find an order-disorder
transition. \textbf{Plot (d)}: The refinement process for the FZ point and
for other 3 additional points: 5-state Potts point, $x_{1}=0.4$, and $%
x_{1}=0.5$. }
\label{Fig:refineent_second_order}
\end{figure*}

For example, by applying our refinement process for $x_{1}=0.42$, the method
produces a clear point where $r$ is maximum $x_{2}^{opt}=0.198(2)$ (see plot
(a) in Fig. \ref{Fig:refineent_second_order}). This value is in complete
agreement with the exact value of the self-dual line, $x_{2}=\frac{(\sqrt{5}%
-1)}{2}-0.42=\allowbreak 0.198\,03...$ . However, it is important to notice
that we did not find the two points which we would expect by looking into
the phase diagram corresponding to the two critical lines. In order to
better exploit such specificities, we simulated our method for two other
inputs: $x_{1}=0.44$ and $x_{1}=0.46$, the first one corresponds to the end
of soft-order transition and the second one was chosen because there is no
ordered phase at this point (see plots (b) and (c) in Fig. \ref%
{Fig:refineent_second_order}).

In those cases we can clearly see that there is no a unique point where $r$
assumes a maximum value. Finally in the same Fig. \ref%
{Fig:refineent_second_order} (plot (d)) we show the behavior of this same
coefficient for some important points just for an appropriated comparison:
the 5-state Potts model (weak first-order transition point), $x_{1}=0.4\ $%
(crossing two second-order lines), $x_{1}=0.5$, and specially the FZ point
whose critical exponents are estimated in this paper. Now we would like to
consider alternatives to determine (localize) points after the bifurcation
point that are localized on the soft-disorder transition line. From now on,
we will be much more empirical in our techniques. As we reported above, our
optimization method captures the points on the self-dual line but the points
corresponding to soft-disorder and soft-order transitions seems to be
neglected by the method and this deserves a better investigation.

Since we used the power laws for ordered initial spin systems, this can be
the reason whereas such transitions are not order-disorder-like. In order to
localize such points we prepared a second algorithm similar to the previous
method. However, instead of optimizing the Eq. \ref{decay_ferro}, by
performing several time-dependent MC simulations starting from $m_{0}=1$, we
monitored simulations starting from $m_{0}=M_{1}(0)=0$ and, in this case, we
expected that the second moment of the order parameter has the power law
given by Eq. \ref{M2} (see \footnote{%
Here it is important to mention that the lattice was randomly vanished by
considering only two spin variables, $n_{i}=1$ and $2$, differently of the
experiments performed to calculate the critical exponents where the lattice
was vanished by putting 1/5 of spin variables of each kind. Our choice was
based on numerical experiments that showed to be appropriated for this kind
of analysis. On the other hand, for the former prepared initial
configurations, the exponent $\zeta $ probably does not correspond to the
correct value $(d-2\beta /\nu )/z$. However, this does not forbid our
approach whereas in this stage of the paper, our aim was only to explore
alternatives for the localization of the critical points and not to estimate
critical exponents which was correctly performed in the appropriate section}%
). Moreover, we also monitored the value of $\varsigma $ whereas it can be
estimated, even without significance, when the coefficient of determination
is not satisfactory.

\begin{figure}[th]
\begin{center}
\includegraphics[width=\columnwidth]{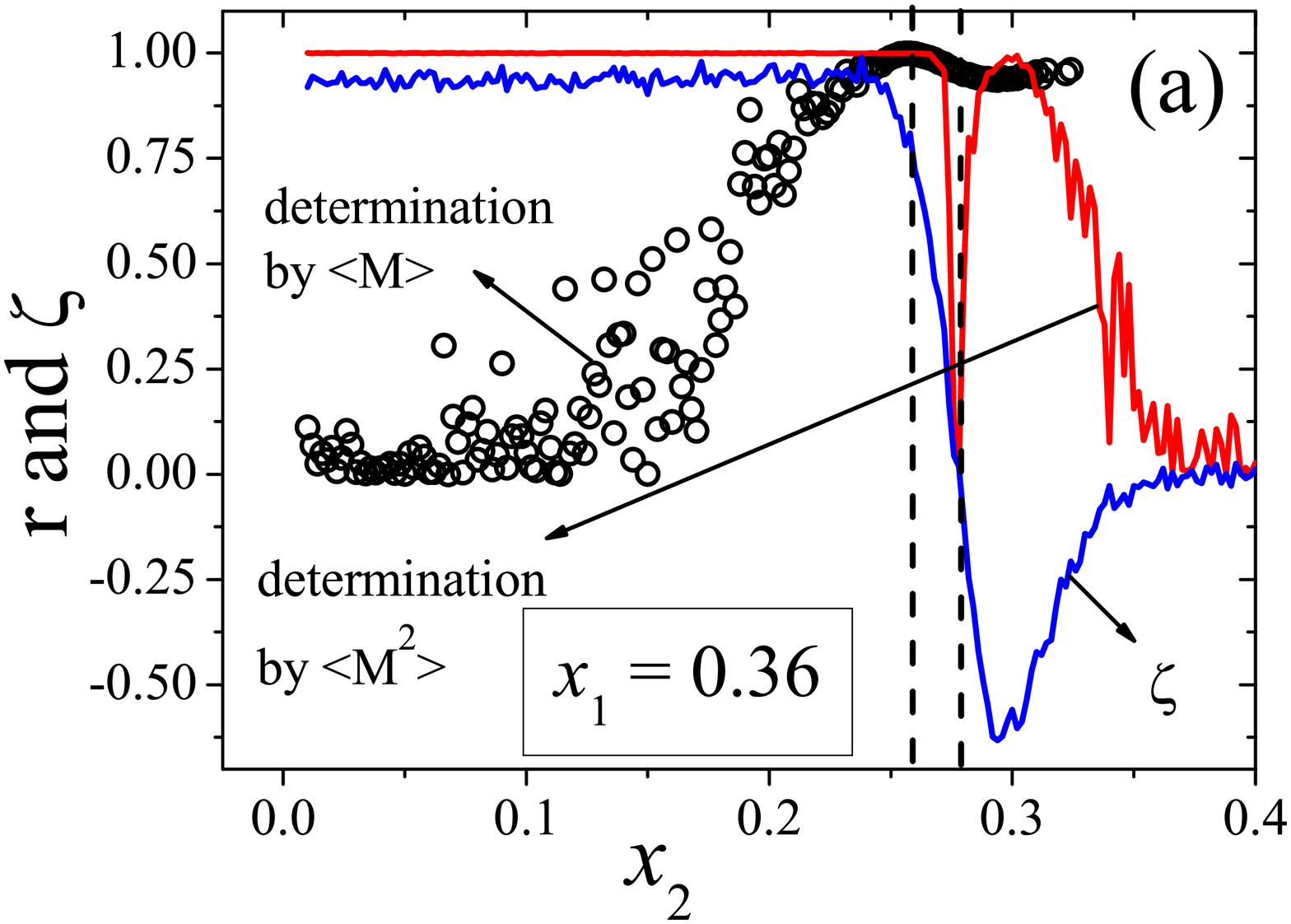} \includegraphics[width=%
\columnwidth]{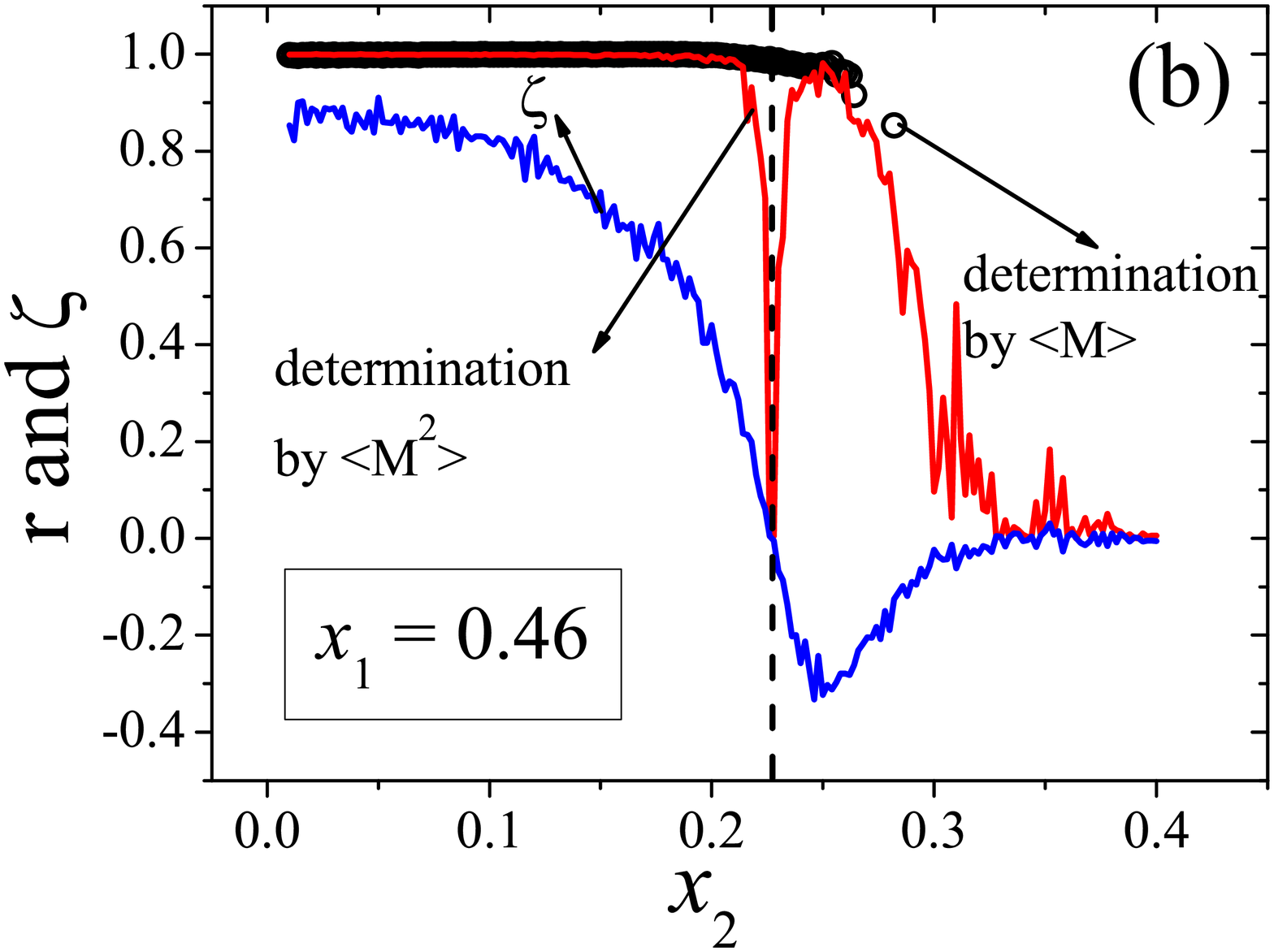}
\end{center}
\caption{(color online) Coefficient of determination for the two different power law fits: $%
\left\langle M\right\rangle _{m_{0}=1}\sim t^{-\protect\beta /\protect\nu z}$
and $\left\langle M^{2}\right\rangle _{m_{0}=0}\sim t^{(d-2\protect\beta /%
\protect\nu )z}$, and evaluation of the coefficient $\protect\varsigma =(d-2%
\protect\beta /\protect\nu )z$ for the different values of $x_{2}$
considering as input: $x_{1}=0.36$ (\textbf{plot a}) and $x_{1}=0.46$ (%
\textbf{plot b})}
\label{M2_monitoring}
\end{figure}

Fig. \ref{M2_monitoring} shows the behavior of the coefficient of
determination when one takes into account the power laws for $\langle
M\rangle _{m_{0}=1}$ and $\langle M^{2}\rangle _{m_{0}=0}$ along with the
numerical estimates of $\varsigma $, for two input values: $x_{1}=0.36$
(plot (a)) and $x_{1}=0.46$ (plot (b)). We can see that determination for $%
\langle M^{2}\rangle _{m_{0}=0}$ for both values decreases abruptly for a
value of $x_{2}$ followed by a subsequent abrupt increase. Such behavior was
found for other several studied points ranging from the 5-state Potts model
to $x_{1}=0.6$. We also can see that the peak of the curves of coefficient
of determination correspond to the points where the numerical estimates of $%
\varsigma $\ change their signal. For instance, for $x_{1}=0.36 $, we found
a clear maximum of the determination coefficient for $x_{2}=0.258(2)$ when
we consider fits for $\langle M\rangle $ (Eq. \ref{decay_ferro}). On the
other hand, when one considers fits for $\langle M^{2}\rangle $ (Eq. \ref{M2}%
) the value of $x_{2}$ at the peak of the determination coefficient ($%
x_{2}=0.278(2)$) does not coincide with the previous one.

In order to establish some relationships between the estimates of the points
where there is an abrupt decreasing of coefficient $r$ for $\langle
M^{2}\rangle $ and values of the soft-disorder transition, we decided to
digitize the phase diagram of the model (Fig. 1, Ref. \cite{Rouidi1992})
in order to localize (by using a pointer on the bitmap figure) and compare
some points of soft-disorder phase transition to the values obtained in our
simulations.

\begin{table}[tbp] \centering%
\begin{tabular}{lll}
\hline\hline
$x_{1}$ & $x_{2}$(Ref. \cite{Rouidi1992}) & $x_{2}$(EM) \\ \hline\hline
FZ & $0.270$ & $0.288(2)$ \\ 
$0.36$ & $0.264$ & $0.278(2)$ \\ 
$0.40$ & $0.252$ & $0.250(2)$ \\ 
$0.44$ & $0.230$ & $0.232(2)$ \\ 
$0.48$ & $0.220$ & $0.220(2)$ \\ 
$0.52$ & $0.209$ & $0.208(2)$ \\ \hline\hline
\end{tabular}%
\caption{Values of $x_{2}$ for several points (first column) obtained through two methods. The second column 
presents the estimates extracted by the digitalization of the Fig. 1 in Ref. \cite{Rouidi1992} and the third column 
shows the values obtained by an alternative empirical method (EM)}\label%
{Table:digitalized}%
\end{table}%

We can observe that after $x_{1}=0.40$ (see TABLE \ref{Table:digitalized})
there is an excellent agreement between unofficial estimates (Ref. \cite%
{Rouidi1992} ) and our empirical method (EM). It is important to mention
that before $x_{1}=0.44$ our method for optimization of the power law for $%
\langle M\rangle _{m_{0}=1}$ has already localized very well the considered
points on the self-dual transition line. So from this analysis we have
two important conclusions:

\begin{enumerate}
\item By taking into account points with $(\sqrt{5}-1)/4<x_{1}<0.44$, we are
able to estimate the best values of $x_{2}$\ which corresponds to the
self-dual line by optimizing the Eq. \ref{decay_ferro}.

\item For $x_{1}\geq 0.40$\ we estimated some values of $x_{2}$\ through the
Eq. \ref{M2} by using an empirical approach and analyzed the soft-disorder
transition, the only transition above the self-dual line, in this region
predicted by the phase diagram (see \cite{Rouidi1992}).
\end{enumerate}

Finally, it is important to mention a technical detail in our simulations.
Here, our initial condition for obtaining $m_{0}=0$ for $\left\langle
M^{2}\right\rangle \ $was built only with spins related to the first order
parameter (Eq. \ref{M_1}), i.e., $n_{i}=1$ or $2$. This case does not
correspond to the correct critical values of $\beta $ and $\nu $, whereas
the correct way to vanish the initial configuration is to put $%
n_{i}=0,1,2,3,4$ in the proportion of 1/5 for each one, as used in this
paper to compute the critical exponents. However, when considering the
empirical method presented above this initial condition ($n_{i}=1$ or $2$)
brings a change of signal of $\varsigma $ which was not observed when
considering the initial natural condition (proportion of 1/5).

\section{Results II: Estimating the critical exponents (Static and dynamic
ones) of the bifurcation point}

\label{Section:ResultsII}

Now we explored the critical exponents of Z(5) model with special attention
to the bifurcation point. Before showing the estimates for this point, we
presented some estimates of the exponent $\mu _{i}=-\beta /\nu z$ from Eq. %
\ref{decay_ferro}, with $i=1$ or $2$, along self-dual line by using the two
order parameters $M_{i}$ (Eqs. \ref{M_1} and \ref{M_2}). Our main idea here
is to study the symmetry between these two order parameters via
non-equilibrium MC simulations and to explore if there is some pair $%
(x_{1},x_{2})$ for which $\mu _{1}=\mu _{2}$. It is important to mention
that $\mu $\ is a sort of effective exponent since it was used to analyze
first weak and second order points. 

\subsection{Exploring the self-dual line}

We prepared an algorithm that measures $\mu $ for each $(x_{1},x_{2})$ pair
in the self-dual line: $x_{2}=(\sqrt{5}-1)/2-x_{1}$ and performed
time-dependent MC\ simulation to obtain averages of the order parameter (Eq. %
\ref{M_1} and \ref{M_2}) and, consequently, the exponents $\mu _{1}$ and $%
\mu _{2}$ from the power law decay (Eq. \ref{decay_ferro}). For these
simulations, we considered $x_{1}$ ranging from $x_{1}^{(\min )}=0.2$ to $%
x_{1}^{(\max )}=0.4$, with $\Delta x_{1}=5\cdot 10^{-3}$. For each input
pair $(x_{1},x_{2})$ we used $N_{run}=1200$ runs, $N_{MC}=150$ and $L=160$
(enough after a fast finite size scaling study as shown in the next
subsection).

\begin{figure}[th]
\begin{center}
\includegraphics[width=\columnwidth]{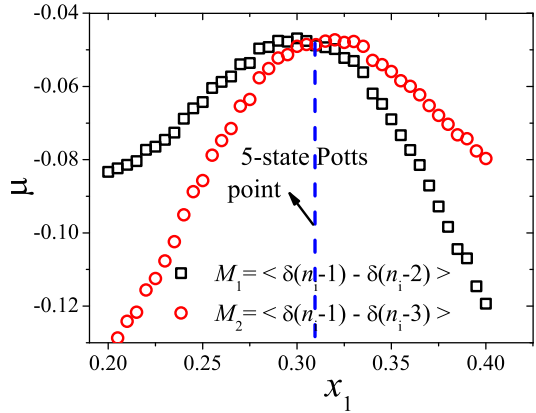}
\end{center}
\caption{(color online) Estimates of the exponent $\protect\mu =-\protect\beta /\protect\nu %
z$ (a sort of effective exponent) along self-dual line. We can observe that
curves assume the same value in $x_{1}=0.310(5)$ which corresponds to the
5-state Potts point.}
\label{Fig:Self_dual_line}
\end{figure}

In Fig. \ref{Fig:Self_dual_line} we show the behavior of $\mu _{1}$ and $\mu
_{2}$ as function of $x_{1}$. We can observe that the curves meet each other
at the point $x_{1}=0.310(5)$ which corresponds to the numerical estimate of
the 5-state Potts point as well as to the symmetry found in the phase
diagram presented in Fig. \ref{Fig:phase_diagram}. Undoubtedly, this is
another interesting finding obtained when using non-equilibrium MC
simulations. It is important to say that we obtained a goodness-of-fit (see
for example \cite{Press1992}) above $0.99$\ for all considered points
showing that all estimates were obtained with robust power law decays. After
these preliminary explorations of the self-dual line and its symmetry via
non-equilibrium MC simulations we explored the numerical estimates of the
critical exponents at the FZ point.

\subsection{The exponents $z$, $\protect\nu $ and $\protect\beta $ of the FZ
point}

\begin{figure}[th]
\begin{center}
\includegraphics[width=\columnwidth]{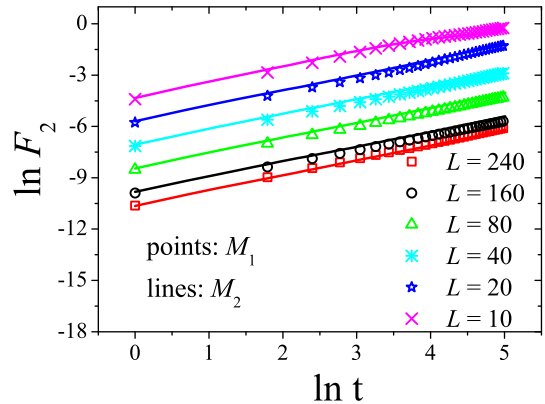} 
\end{center}
\caption{(color online) Time evolution of $F_{2}$ in a ln-ln plot. The points correspond to
the behavior of order parameter $M_{1}$ while lines correspond to the order
parameter $M_{2}$. }
\label{F2}
\end{figure}

Initially we performed simulations to obtain $F_{2}$ as function of $t$. In
order to verify the finite-size effects, we have used lattice of linear
sizes, $L=10$, $20$, $40$, $80$, $160$, and $240$. In Fig. \ref{F2} we can
observe robust power laws for the time evolution of the ratio $F_{2}$. As
can be seen in the figure, the power law behavior of the first order
parameter, $M_{1}$, is showed as point while the second one, $M_{2}$, is
represented by lines. Then, it is possible to notice in this figure that
both order parameters share the same exponent $z$.

In our experiments we used $N_{MC}=150$ MC steps and calculated the
exponents for different time windows of size $\Delta N=10$ MC steps with
respective goodness of fit $q$. In TABLE \ref{Table_M1} (3rd column) we show
the different values obtained for $z$. All intervals presented excellent
goodness of fit (6rd column), with $q_{z}>0.73$.

\begin{table}[tbp] \centering%
\begin{tabular}{ccccccc}
\hline\hline
Interval & $\phi =1/\nu z$ & $z$ & $\mu =\beta /\nu z$ & $q_{1/\nu z}$ & $%
q_{z}$ & $q_{\beta /\nu z}$ \\ \hline\hline
\multicolumn{1}{l}{$\lbrack 30,40]$} & \multicolumn{1}{l}{$0.666(6)$} & 
\multicolumn{1}{l}{$2.38(3)$} & \multicolumn{1}{l}{$0.0641(2)$} & 
\multicolumn{1}{l}{$0.994$} & \multicolumn{1}{l}{$0.998$} & 
\multicolumn{1}{l}{$0.989$} \\ 
\multicolumn{1}{l}{$\lbrack 40,50]$} & \multicolumn{1}{l}{$0.649(5)$} & 
\multicolumn{1}{l}{$2.43(5)$} & \multicolumn{1}{l}{$0.0650(4)$} & 
\multicolumn{1}{l}{$0.998$} & \multicolumn{1}{l}{$1.000$} & 
\multicolumn{1}{l}{$1.000$} \\ 
\multicolumn{1}{l}{$\lbrack 50,60]$} & \multicolumn{1}{l}{$0.667(6)$} & 
\multicolumn{1}{l}{$2.34(6)$} & \multicolumn{1}{l}{$0.0650(7)$} & 
\multicolumn{1}{l}{$0.999$} & \multicolumn{1}{l}{$1.000$} & 
\multicolumn{1}{l}{$1.000$} \\ 
\multicolumn{1}{l}{$\lbrack 60,70]$} & \multicolumn{1}{l}{$0.659\left(
6\right) $} & \multicolumn{1}{l}{$2.40(5)$} & \multicolumn{1}{l}{$0.065(1)$}
& \multicolumn{1}{l}{$0.995$} & \multicolumn{1}{l}{$1.000$} & 
\multicolumn{1}{l}{$1.000$} \\ 
\multicolumn{1}{l}{$\lbrack 70,80]$} & \multicolumn{1}{l}{$0.64(1)$} & 
\multicolumn{1}{l}{$2.28(6)$} & \multicolumn{1}{l}{$0.066(1)$} & 
\multicolumn{1}{l}{$1.000$} & \multicolumn{1}{l}{$1.000$} & 
\multicolumn{1}{l}{$1.000$} \\ 
\multicolumn{1}{l}{$\lbrack 80,90]$} & \multicolumn{1}{l}{$0.66(2)$} & 
\multicolumn{1}{l}{$2.24(6)$} & \multicolumn{1}{l}{$0.066(1)$} & 
\multicolumn{1}{l}{$1.000$} & \multicolumn{1}{l}{$1.000$} & 
\multicolumn{1}{l}{$1.000$} \\ 
\multicolumn{1}{l}{$\lbrack 90,100]$} & \multicolumn{1}{l}{$0.65(2)$} & 
\multicolumn{1}{l}{$2.34(6)$} & \multicolumn{1}{l}{$0.067(2)$} & 
\multicolumn{1}{l}{$0.998$} & \multicolumn{1}{l}{$1.000$} & 
\multicolumn{1}{l}{$1.000$} \\ 
\multicolumn{1}{l}{$\lbrack 100,110]$} & \multicolumn{1}{l}{$0.63(2)$} & 
\multicolumn{1}{l}{$2.35(5)$} & \multicolumn{1}{l}{$0.065(1)$} & 
\multicolumn{1}{l}{$1.000$} & \multicolumn{1}{l}{$0.993$} & 
\multicolumn{1}{l}{$1.000$} \\ 
\multicolumn{1}{l}{$\lbrack 110,120]$} & \multicolumn{1}{l}{$0.66(1)$} & 
\multicolumn{1}{l}{$2.32(3)$} & \multicolumn{1}{l}{$0.067(2)$} & 
\multicolumn{1}{l}{$0.999$} & \multicolumn{1}{l}{$0.917$} & 
\multicolumn{1}{l}{$1.000$} \\ 
\multicolumn{1}{l}{$\lbrack 120,130]$} & \multicolumn{1}{l}{$0.64(2)$} & 
\multicolumn{1}{l}{$2.32(4)$} & \multicolumn{1}{l}{$0.066(3)$} & 
\multicolumn{1}{l}{$1.000$} & \multicolumn{1}{l}{$0.968$} & 
\multicolumn{1}{l}{$1.000$} \\ 
\multicolumn{1}{l}{$\lbrack 130,140]$} & \multicolumn{1}{l}{$0.68(2)$} & 
\multicolumn{1}{l}{$2.33(5)$} & \multicolumn{1}{l}{$0.066(2)$} & 
\multicolumn{1}{l}{$1.000$} & \multicolumn{1}{l}{$0.986$} & 
\multicolumn{1}{l}{$0.999$} \\ 
\multicolumn{1}{l}{$\lbrack 140,150]$} & \multicolumn{1}{l}{$0.66(1)$} & 
\multicolumn{1}{l}{$2.29(4)$} & \multicolumn{1}{l}{$0.067(3)$} & 
\multicolumn{1}{l}{$0.999$} & \multicolumn{1}{l}{$0.737$} & 
\multicolumn{1}{l}{$1.000$} \\ \hline\hline
\end{tabular}%
\caption{Estimates of exponents for different time windows by using the
order parameter $M_{1}$}\label{Table_M1}%
\end{table}%

\begin{figure}[th]
\begin{center}
\includegraphics[width=\columnwidth]{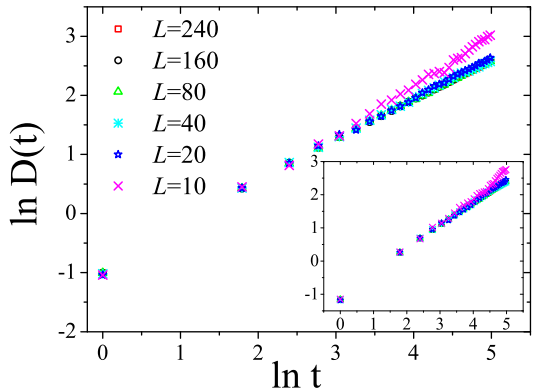} 
\end{center}
\caption{(color online) Time evolution of $D(t)$ in a ln-ln plot for order parameter $M_{1}$%
. The inset plot represents the same time evolution for the order parameter $%
M_{2}$. Just for $L=10$ we can observe a visual reasonable deviation of the
power law behavior. }
\label{Fig_derivada}
\end{figure}

\begin{figure}[th]
\begin{center}
\includegraphics[width=\columnwidth]{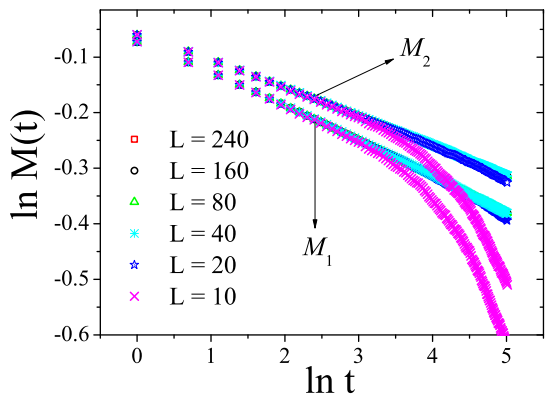} 
\end{center}
\caption{(color online) Decay of the magnetization starting from an ordered initial state.
The branches for each order parameter, $M_{1}$ and $M_{2}$, are indicated in
plot. The difference between the slopes indicates the difference between
critical exponents $\protect\beta _{1}$ and $\protect\beta _{2}$. }
\label{Fig_magordered}
\end{figure}

Similarly, the plots in Figs. \ref{Fig_derivada} and \ref{Fig_magordered}
show the time evolution of $D(t)$ and $M(t)$, for the two different order
parameters. 
Here, $D(t)$ was numerically estimated according to 
\[
D(t)\approx \frac{1}{2\delta }\ln \left[ \frac{\left\langle M\right\rangle
(t,T_{c}+\delta )}{\left\langle M\right\rangle (t,T_{c}-\delta )}\right] 
\]%
where $\left\langle M\right\rangle (t,T_{c}\pm \delta )$ means the
magnetizations above (below) critical temperature of a quantity $\delta $,
starting from ordered initial state. Since our parameters are $%
k_{1}=J_{1}/k_{B}T$ and $k_{2}=J_{2}/k_{B}T$ a perturbation of $\delta $ in $%
T$ corresponds to $k_{1}^{\prime }=J_{1}/k_{B}(T\pm \delta )=k_{1}/(1\pm
\delta ^{\prime })$ and $k_{2}^{\prime }=k_{2}/(1\pm \delta ^{\prime })$,
where $\delta ^{\prime }=\delta /T$.

In TABLE \ref{Table_M1} we also present our results for $\phi $ and $\mu $
exactly as previously reported for $z$. We can observe again good fits in
all time windows. All the analysis and estimates presented above for $M_{1}$
were also performed for the second order parameter, $M_{2}$. However, for
economy they were not reported here whereas a compilation of our main
estimates, including $M_{1}$ and $M_{2}$ are presented in \ref{table:final}.
The results from 2nd to 7th columns are estimated by using the regular
method to obtain the error bars in the context of short time critical MC
simulations, via error propagation (see first part of the appendix).

In this table, the term "best" means the best value found which reproduces
the most similar conjectured values for the static exponents $\nu $ and $%
\beta $ (10th and 11th columns, respectively). The term "prop" refers to
uncertainty which was calculated by error propagation. The term "aver" means
the average of exponents performed from larger time windows taking the
estimates from $[70,80]$ up to $[140,150]$.

\begin{table*}[tbp] \centering%
\begin{tabular}{lllllllllll}
\hline\hline
OP & $\nu _{best}^{(prop)}$ & $\nu _{aver}^{(prop)}$ & $\beta
_{best}^{(prop)}$ & $\beta _{aver}^{(prop)}$ & $z_{best}^{(prop)}$ & $%
z_{aver}^{(prop)}$ & $\nu _{best}^{(boot)}$ & $\beta _{best}^{(boot)}$ & $%
\nu _{exact}$ & $\beta _{exact}$ \\ \hline\hline
$M_{1}$ & $0.70(2)$ & $0.66(1)$ & $0.107(4)$ & $0.105(3)$ & $2.28(6)$ & $%
2.31(1)$ & $0.70(3)$ & $0.119(3)$ & $0.7$ & $0.12$ \\ 
$M_{2}$ & $0.70(3)$ & $0.68(1)$ & $0.080(1)$ & $0.081(1)$ & $2.28(8)$ & $%
2.26(1)$ & $0.70(4)$ & $0.080(2)$ & $0.7$ & $0.08$ \\ \hline\hline
\end{tabular}%
\caption{Final estimates of critical exponents for both order parameters (OP). Here the "best" denotes the value 
used to obtain the static critical exponents more similar to literature. "aver" denotes the value found by 
performing an average over time windows as shown in table \ref{Table_M1} for the order parameter $M_{1}$}%
\label{table:final}%
\end{table*}%

We used an alternative method to obtain better estimates, considering
bootstrap re-sampling method for the uncertainty calculation (see second
part of the appendix for detailed description). The idea is to overcome
possible statistical correlation among the exponents. The results are
presented in 8th and 9th columns. Our estimates by using bootstrap
re-sampling (boot in table \ref{table:final}) corroborate the exact values
for $\nu $ and $\beta $.

First of all, it is important to mention that we obtained estimates of
exponent $z$ for both order parameters which, to our knowledge, have never
been calculated. We can see values greater than estimates for the Ising
model for example ($2.14\lesssim z\lesssim 2.16$) and 3-state Potts model ($%
z\approx 2.19$) \cite{Silva20021}, but similar to results obtained for the
4-state Potts model ($z\approx 2.29$) \cite{Silva2004}. The exponents $z$,
for both order parameters, are in complete agreement according$\ $to error
bars. By using error propagation, our estimates for $\beta $ ($\beta
^{(prop)}$) over any criteria are rigorously according to conjecture value $%
\beta =0.08$ for the order parameter $M_{2}$. On the contrary, although we
have reasonable results for the order parameter $M_{1}$, $\beta
_{best}^{(prop)}=0.107(4)$ and $\beta _{aver}^{(prop)}=0.105(3)$, the error
bars are not enough to cover the conjectured value $\beta =0.12$.

Alternatively, with the procedure described in the second part of the
appendix that combines bootstrap and selection, we have as best estimate $%
\beta _{best}^{(boot)}=0.119(3)$ satisfying the conjecture.

We finally found $\nu _{best}^{(prop)}=0.70(2)$ and $0.70(3)\ $for $M_{1}$
and $M_{2}$ respectively, which corroborates the conjecture $\nu =0.7$.

\section{DISCUSSION AND CONCLUSIONS}

\label{section4}

In this paper we studied the phase diagram of Z(5)\ model through
non-equilibrium finite size scaling study in the context of time-dependent
MC simulations. We determined some critical values and weak first-order
transition values along the self-dual line with special attention to FZ
point that, to our knowledge, have never been analyzed using this approach.
We also determined some transition points along the soft-disorder transition
line by using a non-conventional way that looks for an abrupt "depression"
on the second moment of the order parameter as function of time. Moreover,
we calculated the exponent $\mu =\beta /\nu z$ for several points on the
self-dual line of the model for the two order parameters and we showed that
these exponents are equal for the two order parameters only for the point
correspondent to the 5-state Potts point.

\section{APPENDIX}

\label{sectionappendix}

In this section we present our methods to estimate uncertainties. In this
paper we used two approaches: (1) error propagation: generally used in short
time dynamics literature and (2) alternative error analysis by using
bootstrap estimate.

\subsection{Error propagation}

\label{sectionerrorpropagation}

In this paper, we used $N_{run}=4\times 10^{5}$\ runs for the computation of
averaged time series of the second moment of the order parameters, Eq. \ref%
{M2}, in which require disordered initial configurations, and $%
N_{run}=10^{4} $\ runs for experiments that demand ordered initial
configurations, such as those which take into account the power laws given
by the Eqs. \ref{decay_ferro}, \ref{z} and \ref{1niz}.

The error bars were obtained from $N_{b}=5$\ different bins. Our results,
presented in the following plots, correspond to more refined estimates $%
\overline{\langle M^{k}(t)\rangle }=(1/N_{b})\sum\nolimits_{i=1}^{N_{b}}%
\langle M^{k}(t)\rangle ^{(i)}$\ and the error bars (standard deviation of
average) were estimated as $\sigma /\sqrt{N_{b}}=\left( \frac{1}{%
N_{b}(N_{b}-1)}\sum\nolimits_{i=1}^{N_{b}}\left[ \langle M^{k}(t)\rangle
^{(i)}-\overline{\langle M^{k}(t)\rangle }\right] ^{2}\right) ^{1/2}$, where 
$\langle M^{k}(t)\rangle ^{(i)}$\ denotes the average of $k$-th moment of
magnetization of the $i$-th bin.

The exponent $z$\ was estimated from Eq. \ref{z} as $\widehat{z}=2/\widehat{%
\xi }$\ (by setting $d=2$) and its error, $\sigma _{z}$, was obtained
through the equation $\sigma _{z}=(2/\widehat{\xi }^{2})\sigma _{\xi }$\ ,
where $\sigma _{\xi }$\ is the error obtained from the power law fit. With
the estimate of $z$\ and its respective uncertanty in hand, we were able to
obtain an estimate of $\nu $\ ($\widehat{\nu }$) through the fitting of the
Eq. \ref{1niz}, i.e., $\widehat{\nu }=\widehat{\phi }^{-1}\widehat{z}^{-1}$,
with its respective uncertainty:

\[
\sigma _{\nu }=\left[ \widehat{\phi }^{-2}\widehat{z}^{-4}\sigma _{z}^{2}+%
\widehat{\phi }^{-4}\widehat{z}^{-2}\sigma _{\widehat{\phi }}^{2}\right]
^{1/2}\text{.} 
\]

Now, we can estimate $\beta $. Whereas we have in hand an estimate of $%
\widehat{\phi }$, we can estimate $\beta $, where by fitting the Eq. \ref%
{decay_ferro} $\widehat{\beta }=$\ $\widehat{\mu }/\widehat{\phi }$, with
respective uncertainty 
\[
\sigma _{\beta }=\left[ \widehat{\phi }^{-2}\sigma _{\mu }^{2}+\widehat{\phi 
}^{-4}\widehat{\mu }^{2}\sigma _{\phi }^{2}\right] ^{1/2}\text{.} 
\]%
\ 

\subsection{ Alternative approach with Bootstrap estimates}

\label{sectionbootstrap}

Now we describe an alternative analysis for estimating exponents with
uncertainties calculated by the bootstrap method. Let us start by the
independent exponent $z$. So, instead of determining this exponent by
combining 5 seeds which corresponds to 5 different time series: $t$\ $\times
F_{2}(t)$, and obtaining the error bars over these 5 seeds for each point of
averaged time series, we used a different procedure. Since we have 5 seeds
for $\left\langle M\right\rangle _{m_{0}=1}$\ and 5 seeds for $\left\langle
M^{2}\right\rangle _{m_{0}=0}$\ we can obtain $N_{bin}=25$\ different time
series $t$\ $\times F_{2}(t)$ by crossing the seeds . So, we obtain $%
N_{sample}^{(boot)}$\ different re-sampled data set obtained with
replacement. For each data set, each time series $[t$\ $\times F_{2}(t)]_{i}$%
\ corresponds to a specific bin $i=1,...,N_{bin}$, and an exponent $z_{i}$\
is calculated. Then, for every re-sampled data set would be for example: $%
sample_{1}=(z_{1}^{(1)},z_{2}^{(1)},...,z_{25}^{(1)})$, $%
sample_{2}=(z_{1}^{(2)},z_{2}^{(2)},...,z_{25}^{(2)})$,..., $%
sample_{N_{sample}}=(z_{1}^{(N_{sample}^{(boot)})},z_{2}^{(N_{sample}^{(boot)})},...,z_{25}^{(N_{sample}^{(boot)})}) 
$. So for every re-sampled data we calculate $\left\langle z\right\rangle
_{(i)}=(z_{1}^{(i)}+...z_{25}^{(i)})/N_{bin}$, and with a sampling
distribution of $\left\langle z\right\rangle _{(i)}$\ we calculate $%
\left\langle z\right\rangle
=(1/N_{sample}^{(boot)})\sum_{i=1}^{N_{sample}^{(boot)}}\left\langle
z\right\rangle _{(i)}$. The standard deviation of the sampling is given by $%
\sigma _{z}=\sqrt{(N_{sample}^{(boot)}-1)^{-1}%
\sum_{i=1}^{N_{sample}^{(boot)}}\left( \left\langle z\right\rangle
_{(i)}-\left\langle z\right\rangle \right) ^{2}}$\ which is a standard error
of the mean (this is the more important point).

Since we obtained previously an estimate of $z$\ , we used it as input and
we calculated $\nu ^{(boot)}\ $by using time series $t$\ $\times \frac{1}{%
2\delta }\ln \left[ \frac{\left\langle M\right\rangle
_{m_{0}=1}(t,k_{c}+\delta )}{\left\langle M\right\rangle
_{m_{0}=1}(t,k_{c}-\delta )}\right] $. We also crossed the seeds to obtain $%
N_{bin}=25$\ bins and for each bin, a linear fit is performed producing $%
\phi _{i}\Longrightarrow \nu _{i}=1/(\phi _{i}\cdot z)$. We repeat the
re-sampling procedure in order to obtain: $\sigma _{\nu }=\sqrt{%
(N_{sample}^{(boot)}-1)^{-1}\sum_{i=1}^{N_{sample}^{(boot)}}\left(
\left\langle \nu \right\rangle _{(i)}-\left\langle \nu \right\rangle \right)
^{2}}$. Finally, since we have estimates for $z$\ and $\nu $\ we repeat the
procedures to obtain the error estimate of $\beta $: a) Linear fits produce $%
\mu _{i}\Longrightarrow \beta _{i}=z\cdot \nu \cdot \mu _{i}$, $%
i=1,...,N_{bin}$; b) Re-sampling to obtain the bootstrap estimate of the
error estimate: $\sigma _{\beta }=\sqrt{(N_{sample}^{(boot)}-1)^{-1}%
\sum_{i=1}^{N_{sample}^{(boot)}}\left( \left\langle \beta \right\rangle
_{(i)}-\left\langle \beta \right\rangle \right) ^{2}}$. The only difference
here is that $N_{bin}=5$\ since there is no crossing of seeds for this
estimate.

So, our method follows the prescription:

\begin{enumerate}
\item We obtain two estimates of the dynamic exponent $z$\ (minimum and
maximum) estimates where the error bars were obtained with bootstrap
re-sampling, under $N_{sample}^{(boot)}=10^{4}$. \ 

\item From these two estimates (input), we obtain a list of worst and best
estimates of the static exponent $\nu $. From these estimates we select the
nearest and the farthest estimates with uncertainties calculated by the
bootstrap method.

\item Finally with best and worst values of $\nu $, our re-sampling
bootstrap results in a list of worst and best estimates of $\beta $\ and its
uncertainty.
\end{enumerate}

For example, for the order parameter $M_{1}$\ we have the results for $z$\
according to 2nd column in TABLE \ref{Table:bootstrapM1} for the different
intervals. Taking the two more different estimates (maximum and minimum) we
replicated the bootstrap method in order to obtain candidate estimates for $%
\nu $\ and $\beta $, which is shown in the columns 3, 4, 5, and last one in
this same table. Here $\nu _{best}$ are the values obtained for $z=$\ $2.25$%
\ while the values for $\nu _{worst}$\ were obtained by using $z=$\ $2.36$\
as input. The columns $\beta _{best}$\ and $\beta _{worst}$\ correspond to
the best and worst values by using previous input values. So we choice $\nu
=0.70(3)$\ and $\beta =0.119(3)$\ as better estimates among best estimates.
Similar analysis was performed for $M_{2}$ which is shown in 8th and 9th
columns in TABLE \ref{table:final} of this manuscript.

\begin{table}[tbp] \centering%
\begin{tabular}{llllll}
\hline\hline
Interval & $z$ & $\nu _{best}$ & $\nu _{worst}$ & $\beta _{best}$ & $\beta
_{worst}$ \\ \hline\hline
$\lbrack 70,80]$ & $2.28(5)$ & $0.71(4)$ & $0.66(4)$ & $0.117(3)$ & $%
0.104(3) $ \\ 
$\lbrack 80,90]$ & $2.25(5)$ & $0.69(1)$ & $0.64(1)$ & $0.117(1)$ & $%
0.104(1) $ \\ 
$\lbrack 90,100]$ & $2.35(4)$ & $0.70(3)$ & $0.65(2)$ & $0.119(3)$ & $%
0.106(3)$ \\ 
$\lbrack 100,110]$ & $2.36(7)$ & $0.71(2)$ & $0.66(2)$ & $0.116(3)$ & $%
0.103(3)$ \\ 
$\lbrack 110,120]$ & $2.32(6)$ & $0.69(2)$ & $0.64(2)$ & $0.117(3)$ & $%
0.104(3)$ \\ 
$\lbrack 120,130]$ & $2.32(6)$ & $0.71(3)$ & $0.66(3)$ & $0.117(3)$ & $%
0.104(2)$ \\ 
$\lbrack 130,140]$ & $2.30(3)$ & $0.68(2)$ & $0.63(2)$ & $0.117(1)$ & $%
0.104(1)$ \\ 
$\lbrack 140,150]$ & $2.27(4)$ & $0.69(2)$ & $0.64(2)$ & $0.118(8)$ & $%
0.105(3)$ \\ \hline\hline
\end{tabular}%
\caption{Results for the bootrap by using the order parameter
$M_{1}$}\label{Table:bootstrapM1}%
\end{table}%
\textbf{\ }

\section*{Acknowledgements}

R. da Silva was partly supported by the Brazilian Research Council CNPq. The
authors thank CESUP (Super Computer Center of Federal University of Rio
Grande do Sul) as well as Professor Leonardo G. Brunet (IF-UFRGS) for the
available computational resources. We are grateful for support from
Clustered Computing (ada.if.ufrgs.br). We also would like to thank for the
anonymous referees of the Physical Review E for helpful suggestions.

\end{document}